\documentclass[twoside,twocolumn,9pt]{article}
\usepackage{extsizes}
\usepackage[super,sort&compress,comma]{natbib} 
\usepackage[version=3]{mhchem}
\usepackage[left=1.5cm, right=1.5cm, top=1.785cm, bottom=2.0cm]{geometry}
\usepackage{balance}
\usepackage{mathptmx}
\usepackage{sectsty}
\usepackage{graphicx} 
\usepackage{lastpage}
\usepackage[format=plain,justification=justified,font={stretch=1.125,small,sf},labelfont=bf,labelsep=space]{caption}
\usepackage{float}
\usepackage{fancyhdr}
\usepackage{fnpos}

\usepackage{algpseudocode}
\usepackage{algorithm}
\usepackage{arevmath}     
\usepackage[english]{babel}
\addto{\captionsenglish}{%
  
}
\usepackage{array}
\usepackage{droidsans}
\usepackage{charter}
\usepackage[T1]{fontenc}
\usepackage[usenames,dvipsnames]{xcolor}
\usepackage{setspace}
\usepackage[compact]{titlesec}
\usepackage{hyperref}

\usepackage{epstopdf}

\definecolor{cream}{RGB}{222,217,201}

\begin{document}

\pagestyle{fancy}
\thispagestyle{plain}
\fancypagestyle{plain}{
\renewcommand{\headrulewidth}{0pt}
}

\makeFNbottom
\makeatletter
\renewcommand\LARGE{\@setfontsize\LARGE{15pt}{17}}
\renewcommand\Large{\@setfontsize\Large{12pt}{14}}
\renewcommand\large{\@setfontsize\large{10pt}{12}}
\renewcommand\footnotesize{\@setfontsize\footnotesize{7pt}{10}}
\makeatother

\renewcommand{\thefootnote}{\fnsymbol{footnote}}
\renewcommand\footnoterule{\vspace*{1pt}%
\color{cream}\hrule width 3.5in height 0.4pt \color{black}\vspace*{5pt}} 
\setcounter{secnumdepth}{5}

\makeatletter 
\renewcommand\@biblabel[1]{#1}            
\renewcommand\@makefntext[1]%
{\noindent\makebox[0pt][r]{\@thefnmark\,}#1}
\makeatother 
\renewcommand{\figurename}{\small{Fig.}~}
\sectionfont{\sffamily\Large}
\subsectionfont{\normalsize}
\subsubsectionfont{\bf}
\setstretch{1.125} 
\setlength{\skip\footins}{0.8cm}
\setlength{\footnotesep}{0.25cm}
\setlength{\jot}{10pt}
\titlespacing*{\section}{0pt}{4pt}{4pt}
\titlespacing*{\subsection}{0pt}{15pt}{1pt}

\fancyfoot{}
\fancyfoot[LO,RE]{\vspace{-7.1pt}\includegraphics[height=9pt]{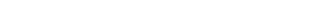}}
\fancyfoot[CO]{\vspace{-7.1pt}\hspace{13.2cm}\includegraphics{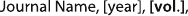}}
\fancyfoot[CE]{\vspace{-7.2pt}\hspace{-14.2cm}\includegraphics{head_foot/RF}}
\fancyfoot[RO]{\footnotesize{\sffamily{1--\pageref{LastPage} ~\textbar  \hspace{2pt}\thepage}}}
\fancyfoot[LE]{\footnotesize{\sffamily{\thepage~\textbar\hspace{3.45cm} 1--\pageref{LastPage}}}}
\fancyhead{}
\renewcommand{\headrulewidth}{0pt} 
\renewcommand{\footrulewidth}{0pt}
\setlength{\arrayrulewidth}{1pt}
\setlength{\columnsep}{6.5mm}
\setlength\bibsep{1pt}

\makeatletter 
\newlength{\figrulesep} 
\setlength{\figrulesep}{0.5\textfloatsep} 

\newcommand{\topfigrule}{\vspace*{-1pt}%
\noindent{\color{cream}\rule[-\figrulesep]{\columnwidth}{1.5pt}} }

\newcommand{\botfigrule}{\vspace*{-2pt}%
\noindent{\color{cream}\rule[\figrulesep]{\columnwidth}{1.5pt}} }

\newcommand{\dblfigrule}{\vspace*{-1pt}%
\noindent{\color{cream}\rule[-\figrulesep]{\textwidth}{1.5pt}} }

\makeatother

\twocolumn[
  \begin{@twocolumnfalse}
{\includegraphics[height=30pt]{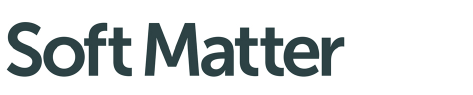}\hfill\raisebox{0pt}[0pt][0pt]{\includegraphics[height=55pt]{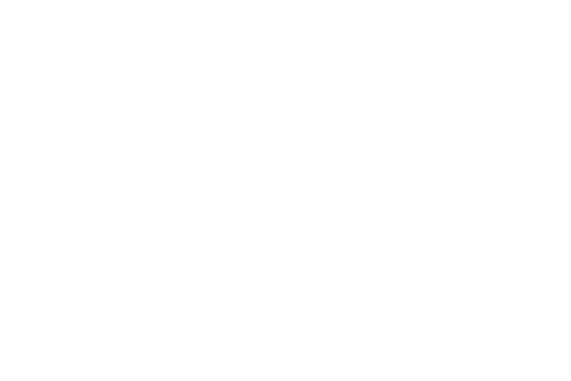}}\\[1ex]
\includegraphics[width=18.5cm]{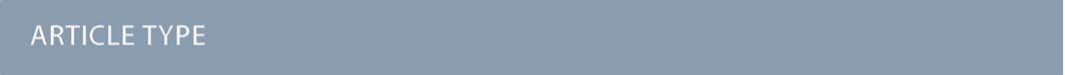}}\par
\vspace{1em}
\sffamily
\begin{tabular}{m{4.5cm} p{13.5cm} }

\includegraphics{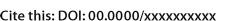} & \noindent\LARGE{\textbf{Tuning the Stability of a Model Quasicrystal and its Approximants with a Periodic Substrate}} \\

\vspace{0.3cm} & \vspace{0.3cm} \\

 & \noindent\large{Nydia Roxana Varela-Rosales\textit{$^{a}$}, Michael Engel$^\dag$\textit{$^{a}$}} \\

\includegraphics{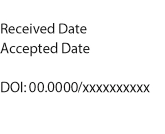} & \noindent\normalsize{
Quasicrystals and their periodic approximants are complex crystalline phases.
They have by now been observed in many metallic alloys, soft matter systems, and particle simulations.
In recent experiments of thin-film perovskites on solid substrates, the type of complex phase was found to change depending on thermodynamic conditions and the type of substrate used.
Here, we investigate the effect of a substrate on the relative thermodynamic stability of a two-dimensional model quasicrystal and its approximants.
Our simulation model are particles interacting via the Lennard-Jones-Gauss potential.
Our numerical methods are molecular dynamics simulations and free energy calculations that take into account phason flips explicitly. 
For substrates interacting weakly with the particles, we observe an incommensurate-commensurate transition, in which a continuous series of quasicrystal approximants locks into a discrete number of approximants.
Interestingly, we observe that the $3/2$ approximant exhibits phason mode fluctuations in thermodynamic equilibrium.
Such fluctuations are reminiscent of a random tiling, a phenomenon usually associated only with quasiperiodic order.
For stronger substrates, we find an enhancement of the stability of the dodecagonal quasicrystal and variants of square lattices.
We explain all observed phenomena by the interplay of the model system with the substrate.
Our results demonstrate that designing novel complex periodic and quasiperiodic structures by choice of suitable substrates is a promising strategy. 
} \\

\end{tabular}

 \end{@twocolumnfalse} \vspace{0.6cm}
  ]

\renewcommand*\rmdefault{bch}\normalfont\upshape
\rmfamily
\section*{}
\vspace{-1cm}


\footnotetext{\textit{$^{a}$Institute for Multiscale Simulation, Friedrich-Alexander-Universität Erlangen-Nürnberg, 91058 Erlangen, Germany. $^{\dag}$E-mail: michael.engel@fau.de}}

\section{Introduction}

The discovery of structures with point group symmetries forbidden in periodic crystals~\cite{Shechtman_1983} created a new field of crystallography, the field of quasicrystals~\cite{Levine_1984,Steurer_2018} (QCs).
Bulk, three-dimensional QCs became popular as candidates for materials with functional properties such as thermal insulators~\cite{thermalConductivity_1996}, coating materials~\cite{coating_pt2_1991,coating_2007}, photonic crystals~\cite{photonic_Vardeny2013} and superconductors~\cite{SuperCond_Kamiya2018}.
The first atomic QC in two dimensions was found only recently in the thin-film perovskite BaTiO$_{3}$ on a Pt substrate~\cite{Nature_Forster2013}.
This discovery inspired follow-up works of thin-film perovskite QCs as well as their approximants at different chemical compositions~\cite{Foerster_OQC_2016,Foerster2017,Foerster_2019,Foerster_2021,Zollner_2020,Zollner_2020_pt2}.
Despite significant efforts, how these perovskite QCs form and why they are stable remain mostly open questions.
Answering these questions is of interest to simulation studies but difficult to address with quantum mechanical methods at the atomic scale because the lack of periodicity and slow phason relaxation in QCs requires large systems over long simulation times~\cite{Mihalkovik_2016,Thiago_2021}.
Coarse-grained model systems are advantageous because they capture the essential physics and offer a route towards elucidating relative thermodynamic stability of the involved phases and their phase transformation pathways.

Two-dimensional particle systems in the presence of a substrate exhibit intricate structure formation phenomena caused by the competition of two length scales.
These length scales are the size of the unit cell of the unperturbed particle system and the periodicity of the substrate.
The competition of the length scales can introduce novel commensurate and incommensurate phases~\cite{monolayer_Patrykiejew2009,sq_sub_Schmiedeberg_2013}.
Incommensurability also plays a role in nanotribology, for example Aubry transitions in systems of colloids sliding on a substrate \cite{experiment_2dAburby_2018,superlubricity_review_2018,Vink_incommensurateSubstrate,Sliding_substrate_2021}.
Three factors are relevant: substrate potential depth, substrate periodicity, and the crystallographic symmetries of the particle system and the substrate~\cite{review_2016_Reichhardt_depinning}.

The investigation of thin QC films on a substrate can be divided into two classes of systems.
The first class comprises systems, which are templated to have QC order by use of a QC substrate.
Such QC films have been realized at the nanoscale as atomic adlayers~\cite{experimental_ex_nanoepitaxy,experimental_ex_LEED,Weisskopf_Y_sim_sub_deca_2007,McGrath_templated_QC_mol_ord_2014}.
And at the colloidal scale, QC laser fields can be finely tuned to induce a gradual transition from an unperturbed system to a frozen QC~\cite{arch_tiles_Mikhael2008,Schmiedeberg_2D_quasicrystSubst,arch_tiles_part2_Schmiedeberg2010,arch_tiles_Mikhael,Jagannathan2014_eight_fold_laser}.
The second class comprises systems with intrinsic QC order.
That is, such systems are QCs even in the absence of the substrate.
It is suspected that thin film perovskite QCs fall into the latter class~\cite{Nature_Forster2013}.
Both classes of systems can show interesting phase behavior as the substrate is tuned.

There exist many simulation models for 2D QCs.
Prominent examples are binary mixtures of particles interacting with the Lennard-Jones potential~\cite{Leung_dod-QC_1989,Fayen2023}, one-component systems of particles with interaction potential minima at two length scales~\cite{Quandt_1999,Engel_2007,Engel_2010}, and anisotropic particles with patchy interactions~\cite{Marjolein_2012,Doye2013}.
Here, we study the Lennard-Jones-Gauss (LJG) potential, which stabilizes a number of one-component QCs and approximants already in the absence of a substrate~\cite{Engel_2007,Engel_F-L_validation}.
We use this coarse-grained model to investigate the possibility of targeting specific approximants by tuning substrate parameters.
As we will see, the LJG system can behave like a thin film QC in both the first and the second class, depending on the choice of thermodynamic conditions and the substrate.

\section{Methods} \label{Methods}
\subsection{Coarse-grained model}

As in our previous work~\cite{Engel_F-L_validation}, we investigate a coarse-grained two-dimensional model system, which is known to form different types of (quasi-)periodic phases~\cite{Engel_2007}.
In this model, identical particles interact isotropically with the LJG potential given by
\begin{equation}
V_\text{LJG}(r) = \frac{1}{r^{12}} - \frac{2}{r^6}  - {\varepsilon}_{LJG} \exp\left( \frac{-{(r-{r_\text{LJG}})}^2 }{2{\sigma_\text{LJG}}^{2} } \right).
\label{eq:LJG}
\end{equation}
The potential parameters are fixed in this work at ${\varepsilon}_\text{LJG} = 1.8$, $r_\text{LJG} = 1.42$, and $\sigma_\text{LJG}^{2} = 0.042$ to stabilize a dodecagonal QC at intermediate temperatures~\cite{Engel_F-L_validation}.
The phase diagram of the LJG potential at these parameters contains a number of phases described by tilings made from squares, triangles, and pentagons.
The sequence of phases can be described according to the underlying symmetry change~\cite{Engel_F-L_validation}
\begin{equation}\label{eq1}
\text{p3m1}\xrightarrow{0.24}\text{p6m}\xrightarrow{0.35}\text{p12m}\xrightarrow{0.36}\text{p4m}\xrightarrow{0.41}E(2),
\end{equation}
where the symbols are the wallpaper groups of the phases and the numbers above the arrows indicates the critical temperature of the phase transitions.
It has been shown that a relatively simple periodic approximant (wallpaper group p3m1) is the energetic ground state stable at $T=0$.
This simple approximant transitions via a series of increasingly more complex hexagonal approximants (wallpaper group p6m) continuously to a dodecagonal QC (wallpaper group p12m) in the temperature range $0.24<T<0.35$.
Towards slightly higher temperature, at $T=0.36$, the QC converts into a square tiling (wallpaper group p4m) before melting at $T=0.41$ into the liquid.
The liquid has no broken rotational or translational symmetry and is thus represented by the group of all isometries, the Euclidean group $E(2)$.

\begin{figure}
    \centering
    \includegraphics[width=1\linewidth]{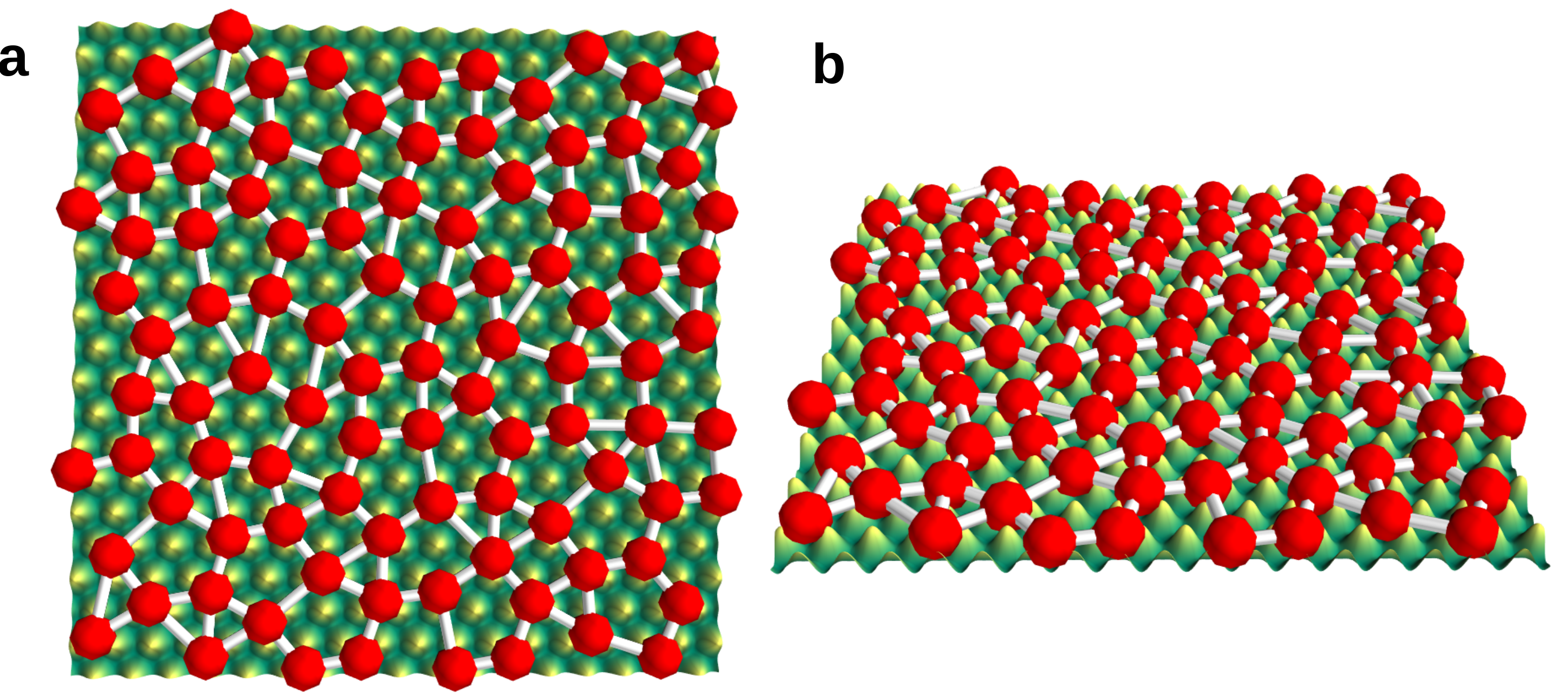}
    \caption{Schematic depiction of the model system of this work.
    Particles (red) interacting with themselves via the LJG potential and with an external potential simulating the substrate.
    The particles form a dodecagonal quasicrystal.
    Bonds connect nearest neighbors and form three types of tiles: triangles, squares, and pentagons.
    Tiles are deformed due to vibrational motion of the particles at an elevated temperature.
    \textbf{a}, Top view. \textbf{b}, Side view at an angled perspective.}
    \label{fig:set_up_visualization}
\end{figure}

The phase behavior of the LJG system is only known in the absence of a substrate.
Here, we add a substrate by including an external potential.
We mimic the effect of the Pt (111) substrate for thin-film perovskite QCs by choosing an external potential in form of a hexagonal lattice by superposition of three plane waves,
\begin{equation}
V_\text{ext} = \varepsilon\frac{\varepsilon_\text{LJG}}{T} \sum_{j=0}^{2}\frac{\cos(2\pi\boldsymbol{k}_{j}\cdot\boldsymbol{r}/\lambda)/3+1}{2}
\end{equation}
with $\boldsymbol{k}_{j} =(\cos(\pi j/3),\sin(\pi j/3))$.
There are two parameters, substrate potential depth $\varepsilon$ and substrate periodicity $\lambda$.
A prefactor of $\varepsilon_\text{LJG}/T$ is included for convenience to ensure that the region of interest is approximately rectangular in the $T$-$\varepsilon$ parameter plane (see assembly diagram below, Fig.~\ref{fig:papermediumpotsimplified}).
A schematic depiction of the model, comprising of the particles (red) interacting with themselves and the periodic substrate is shown in Fig.~\ref{fig:set_up_visualization}.

To justify the choice of model system for this work, we need to discuss phason modes.
Phason modes are dynamic modes characteristic for QCs.
They are realized as collective rearrangements of finite clusters of tiles and can be decomposed into a series of elementary particle displacements called phason flips.
In many common QCs, like for example the Penrose tiling, phason mode excitation requires only low activation energy and can occur locally with only a small number phason flips.
We call phason modes such as those found in the Penrose tiling \emph{continuously excitable}.
In other QCs, for example the square-triangle tiling, phason mode excitation requires creating a pair of defects and propagating each defect along a path rearranging the tiling in the process until the defects annihilate.
In the square-triangle tiling this process is called a zipper.~\cite{Henley1993}
Zippers are not efficient in relaxing the square triangle-tiling.
We call phason modes such as those found in the square-triangle tiling \emph{not continuously excitable}.
Continuously excitable phason modes are found in thin-film perovskite QCs where they involve rhomb tiles with small interior angle $30^\circ$.
Because of the presence of continuously excitable phason modes, perovskite QCs can efficiently relax their tiling and transform to and from approximants~\cite{Foerster_OQC_2016,Foerster2017}.
We chose to study the LJG model system because it stabilizes includes continuously excitable phason modes in the p6m and p12m phases~\cite{Engel_F-L_validation}.
This means our model system is similar to perovskite QCs in the character of its phason flips and the phase transformations between approximants and the QC.

\subsection{Tiling construction in hyperspace}

The tilings of the dodecagonal QC and its approximants in the LJG model system can be described by the cut-and-project method \cite{deBruijn1981,deBruijn1981pt2}
in a four-dimensional hyperspace. We construct tiling models in two steps.

In a first construction step, approximants of the quasicrystalline shield tiling~\cite{Gahler1988} are constructed.
The projection matrices on the two-dimensional physical (or parallel) space is
\begin{equation}
    \boldsymbol{P}^{\parallel} = \frac{1}{\sqrt{2}}
    \begin{bmatrix}     
        c_0&c_1&c_2&c_3\\
        s_0&s_1&s_2&s_3\\
    \end{bmatrix}
\end{equation}
with $c_i=\cos((i+1/2)\pi/6)$ and $s_i=\sin((i+1/2)\pi/6)$.
The projection matrix on the two-dimensional internal (perpendicular) space is
\begin{equation}
    \boldsymbol{P}^{\perp} = \frac{1}{\sqrt{2}} 
    \begin{bmatrix} 
        c'_0&c'_5&c'_{10}&c'_3\\
        s'_0&s'_5&s'_{10}&s'_3\\
    \end{bmatrix}
\end{equation}
with $c'_i=\cos(i\pi/6+\phi)$ and $s'_i=\sin(i\pi/6+\phi)$.
The occupation domain for the shield tiling is a dodecagon with twelve vertices $\boldsymbol{v}_i=(c'_i,s'_i)$, $i=1, \ldots, 12$.
The parameter
\begin{equation}
    \phi = \arctan\left(\frac{t - \sqrt{3}}{t + \sqrt{3}}\right)\quad \text{with}\quad t = 2 \frac{q}{p} - 1
\end{equation}
constructs the dodecagonal shield tiling for $\phi = 0$ and $q/p$ periodic approximants for $\phi \neq 0$.
In this notation, approximants are labeled by two integers, $q$ and $p$.
Their ratio ranges from $q/p = 2$ for the simple periodic approximant with wallpaper group p3m1 to $q/p = (\sqrt{3} + 1) / 2$ for the quasicrystal with wallpaper group p12m.
This means the simple periodic approximant is identified as the $2/1$ approximant.
All approximants with wallpaper group p6m can be constructed in the range $(\sqrt{3} + 1) / 2<q/p<2$.
The approximants have rectangular unit cells with lattice constants $b_x = b_y \sqrt{3}$ and $b_y = (t + \sqrt{3}) p / \sqrt{2}$.
The approximants of the shield tiling constructed so far are built from three basic tiles: square, triangle, and shield\cite{Gahler1988}.

In a second construction step, the basic tiles square, triangle, and shield are decorated by placing one particle in the center of the square tiles, one particle in the center of the triangle tiles, and three particles as a regular triangle in the center of the shield tiles.
The result are the tilings of interest, built from three larger tile types: square, triangle, and slightly deformed pentagon.
Examples of these resultant tilings are found in Fig.~\ref{fig:set_up_visualization} and below in Fig.~\ref{fig:snapshots_examples_stabilityDiagram}.

\subsection{Simulations and free energy calculations}

We perform simulations and free energy calculations using a combination of molecular dynamics (MD) and Monte Carlo (MC) to accelerate relaxation and exploration of phase space.
MD is used to sample phonon modes and configurational entropy.
MC is used to sample phason modes~\cite{Kiselev_2012}.
From the van Hove autocorrelation function~\cite{Engel_2010}, we observe that all phason flips readily occur over a single flip distance near $r=1$ and that most particles can perform phason flips.
This demonstrates that phason modes are continuously excitable via local phason flips and do not require significant activation energy.
We use the HOOMD-blue molecular dynamics simulation package~\cite{hoomd_2015,hoomd_2008} as a basis for our simulations and further extend it to include MC moves.
For MD, the integration timestep is 0.01.
The $NVT$ ensemble with Nosé–Hoover thermostat is used and open boundary conditions are applied.
The Boltzmann constant is set to 1.
For MC, we attempt a phason flip in MC by choosing a particle at random and a flip vector uniformly distributed in the ring $0.9 < r < 1.1$.
The MC move is accepted according to the Metropolis acceptance criterion.
The measured acceptance rate of MC phason flips is about 10\%.
Periodic boundary conditions are employed to minimize finite size effects.
We now discuss the two different simulation modes used in this work.

In the first simulation mode, used for generating a diagram summarizing assembly behavior as a function of model parameters, called an \emph{assembly diagram}, and for studying phase transitions, we perform MD simulations without MC moves.
We simulate 10000 particles in a quadratic simulation box of box edge length 300.
Particles are initially placed in a central circle with number density 0.3.
The particles crystallize over time into a roughly circular, compact cluster, which corresponds to a solid-gas coexistence.
We chose this setup to effectively obtain open boundaries for the crystalline cluster.
Such open boundaries are essential for efficient phason relaxation because phason relaxations are generally suppressed by the topological constraints of periodic boundary conditions.

In the second simulation mode, used for calculation of Helmholtz free energies, we combine MD and MC.
5000 MD sweeps, i.e., moving each particle 5000 times, are alternated with 5000 MC steps, i.e., attempting to perform a phason flip for 5000 randomly chosen particles.
System size varies for different approximants and the QC between 4000 and 12000 particles.
In contrast to the first simulation mode, the simulation box for approximants is now completely filled to prevent a solid-gas coexistence and to avoid phase transitions between approximants.
For the QC, we simulate with open boundaries and only consider particles sufficiently fare from the surface.
We first relax constructed approximants in an $NPT$ simulation and then perform free energy integration in the $NVT$ ensemble.
In this way, the number density of simulations in the second simulation mode is guaranteed to agree with that of simulations in the first simulation mode.

We use the Frenkel-Ladd (FL) method~\cite{Frenkel_1984} to compute free energies.
FL performs a thermodynamic integration over a parameter $\lambda$.
It interpolates between the system of interest at $\lambda=0$ and an Einstein crystal at $\lambda=1$.
In an Einstein crystal, each particle is connected to a reference position by a harmonic spring.
The integration is performed as
\begin{equation}
    F=F_\text{Ein}-\int_0^1 \Delta F(\lambda)\,d\lambda,
\end{equation}
where $\Delta F$ is measured from simulation.
Because the reference free energy of the Einstein crystal, $F_\text{Ein}$, is known, we can calculate in this way absolute free energies.

\begin{figure}
    \centering
    \includegraphics[width=1\linewidth]{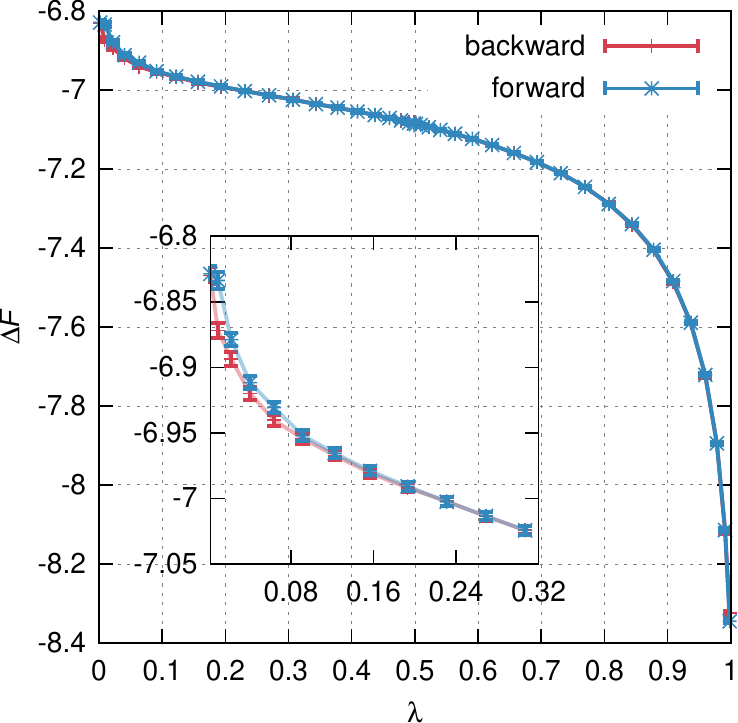}
    \caption{Frenkel-Ladd integration for the $2/1$ approximant at $T=0.25$.
    We measure the integrand $\Delta F$ in simulation by time-averaging at each $\lambda$ point in both forward direction and backward direction.
    The inset shows a zoom-in to small $\lambda$ values.}
    \label{fig:deltaF_approximant_1_2}
\end{figure}

It is important to check that the system remains in equilibrium throughout the FL integration.
We perform such a check by calculating free energy in forward direction (increasing $\lambda$) and in backward direction (decreasing $\lambda$) and by checking for hysteresis. Absence of hysteresis is an indication that free energy calculation is accurate. A test FL integration for the $2/1$ approximant is shown in Fig.~\ref{fig:deltaF_approximant_1_2}.
Forward and backward curves for $\Delta F$ agree, except for a small difference near $\lambda=0$.
At small $\lambda$, particles diffuse strongly through the system, which slows down equilibration.
We use our test FL integration to adjust the numerical parameter to ensure that hysteresis is sufficient small. For numerical integration in production calculations, we use Legendre-Gauss quadrature with 20 $\lambda$ points. Each $\lambda$ point is simulated for 25000 MD sweeps.
The spring constant of the Einstein crystal is $\Lambda_\text{E}=100$.
This value guarantees similar vibrational dynamics in the Einstein crystal and the system of interest as indicated by an approximately flat integrand~\cite{Punnathanam2021}.

We validate the FL implementation by comparison to reported free energies~\cite{Engel_F-L_validation} for the continuous transition $\text{p3m1}\rightarrow\text{p12m}$ in the absence of a substrate at $\varepsilon=0$.
We calculate free energies $F_n(T)$ for six constructed approximants $q/p=n/10$ with $n\in[15,\ldots,20]$ and a well-equilibrated QC.
We follow the continuous transition from the $2/1$ approximant to the quasicrystal via the superstructure wave vector~\cite{Engel_F-L_validation}
\begin{equation}\label{eqn:qs}
    q_{s} = \frac{8\pi}{\sqrt{6}}\frac{2p-q}{2q+(\sqrt{3}-1)p}.
\end{equation}
This wave vector is calculated from the distance between first-order satellite peaks and main diffraction peaks in reciprocal space~\cite{Engel_F-L_validation}.
The appearance of satellite peaks is a consequence of the presence of structural modulations in the approximants.
We define the stability temperature of the approximant $n/10$ as the intersection of $F_{n-1}$ and $F_{n+1}$.
As an exception, instead of the $14/10$ approximant, we use a well-equilibrated quasicrystal.

\begin{figure}
    \centering
    \includegraphics[width=1\linewidth]{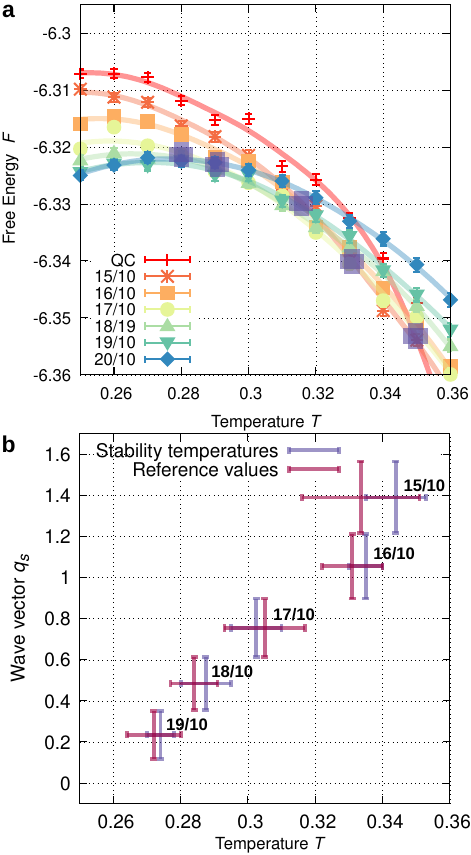}
    \caption{\textbf{a}, Helmholtz free energies $F_n(T)$ for a well-equilibrated quasicrystal and six approximants $n/10$, $n=15,\ldots,20$. Standard errors are estimated from eight independent FL calculations. Purple plus markers indicate the stability temperatures of approximants.
    Data shows the forward branch of the free energy calculation. \textbf{b},~Validation of the free energy calculation with the FL plus phason model relaxation method. The superstructure vector grows linearly with temperature from the $2/1$ approximant to the QC. Error bars on the temperature axis connect the FL calculation for the forward and the backward calculation. Stability temperatures for approximants agree with reference values~\cite{Engel_F-L_validation} within the error bars.}
    \label{fig:fe_qc_vs_2}
\end{figure}

Our implementation of the FL plus phason mode relaxation algorithm shown in Fig.~\ref{fig:fe_qc_vs_2}a accurately reproduces prior work~\cite{Engel_F-L_validation}.
In particular, we reproduce the reported linear increase of the wave vector $q_s$ as a function of temperature (Fig.~\ref{fig:fe_qc_vs_2}b).
The agreement with prior work validates our free energy calculation with the FL plus phason model relaxation method.

\section{Results}   \label{Results}
\subsection{Molecular dynamics simulations}

We analyze the crystallization behavior of our system by performing 77 MD simulations at seven temperatures $0.2\le T\le0.5$ in steps of $0.05$ and for eleven substrate potential depths $0.0\le\varepsilon\le0.1$ in steps of $0.01$.
Substrate periodicity is fixed at $\lambda=0.5$.
This value is inspired by a direct comparison of relevant length scales in our simulations and the perovskite QCs on a Pt substrate~\cite{Foerster2017}.
Each simulation is run for $10^8$ integration steps and the final simulation snapshot at the end of the run is analyzed.
In this subsection, we first perform a general analysis of the orientational order and then conduct a more in-depth analysis of the observed structures by visual inspection and with the help of diffraction patterns.

\subsubsection{Local orientational order analysis}

We quantify the presence of $n$-fold local orientational order via the Mermin order parameter~\cite{Mermin_1968} defined as 
\begin{equation}
    \Psi_{n} = \frac{1}{N} \sum_j \frac{1}{N_j} \sum_{k\text{ NN } j} \exp (in\theta_{jk}),
    \label{eq:mermin}
\end{equation}
where $N$ is the number of particles, $N_j$ the number of nearest neighbors of particle $j$ using the cut-off $r=2.5$, and $\theta_{jk}$ the orientation of the nearest neighbor bond connecting particle $j$ and particle $k$.
The Mermin order parameters for $n=6$ and $n=4$ are also called the hexatic and quadratic order parameter, respectively.
We are specifically interested in these order parameters because we expect to observe a competition between the quadratic order at high temperature $T$ (as previously found in the LJG potential~\cite{Engel_F-L_validation}) and the hexagonal order of the substrate.

\begin{figure}
    \centering
    \includegraphics[width=1\linewidth]{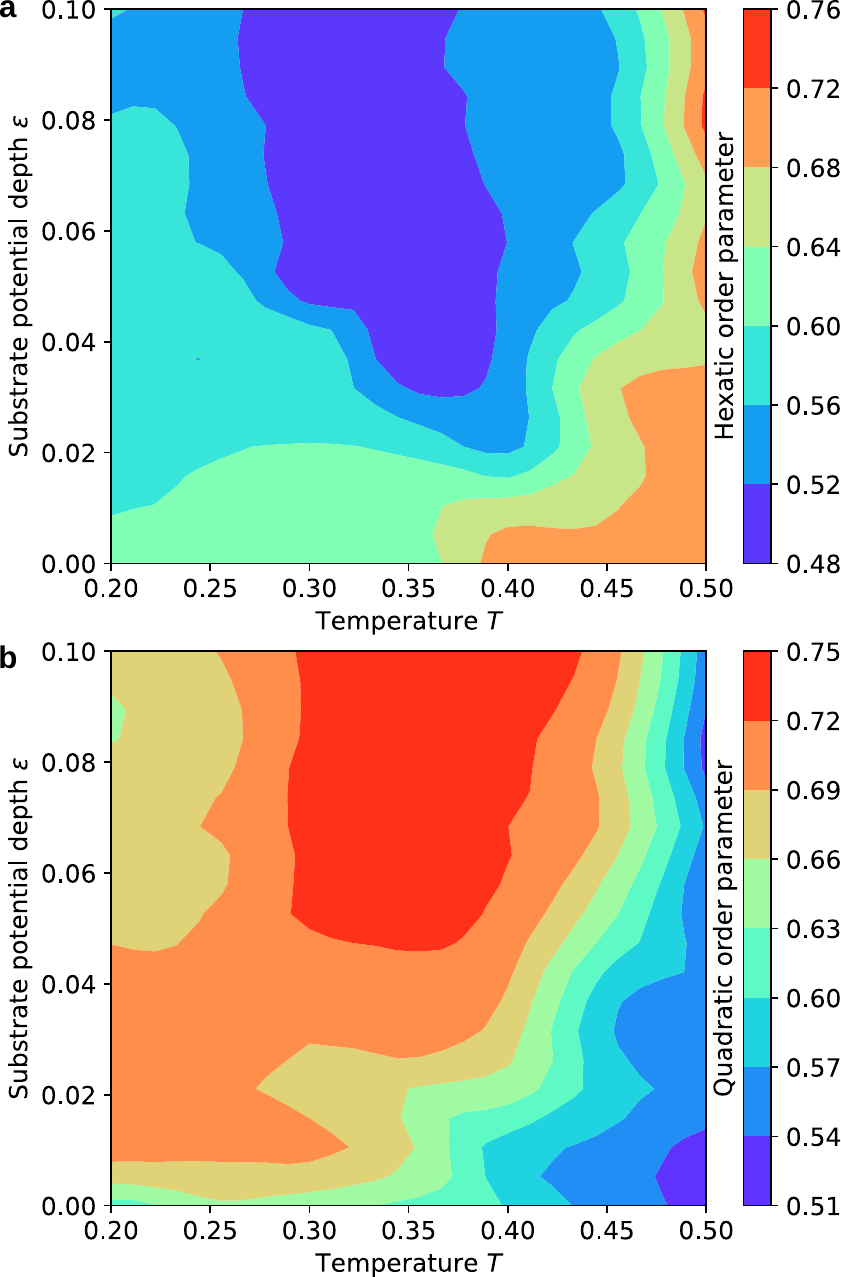}
    \caption{Prevalence of local orientational order in final simulation snapshots of long MD simulations. Data is smoothed by cubic interpolation.
    \textbf{a}.~Hexatic order parameter $\Psi_{6}$.
    \textbf{b}.~Quadratic order parameter $ \Psi_{4}$.}
	\label{fig:compendium_hexatic_quadratic}
\end{figure}

The results of the local orientational order analysis are shown in Fig.~\ref{fig:compendium_hexatic_quadratic}.
Indeed, there exists a competition between quadratic and hexagonal local orientational order.
Hexagonal local order dominates at high $T$.
Here the liquid phase is expected.
In contrast, quadratic local order dominates at intermediate $T$ and high $\varepsilon$.
Apparently, the substrate does not promote local hexagonal order but instead local quadratic order.
This observation will be discussed further below.
The gradual variation of local orientational order across the $T$-$\varepsilon$ parameter plane indicates that the particles can react smoothly to changes of temperature and the substrate at a local level.
We also observe that the presence of thermal fluctuations dominates over crystallographic symmetry in our two-dimensional system, smearing out the boundaries separating ordered phases.
This precludes using the Mermin order parameters for automatic calculation of a assembly diagram without relaxation or temporal averaging.

\subsubsection{Classification of observed phases}

To obtain an assembly diagram, we analyze each final simulation snapshot in direct space by inspection of the particle configuration as well as in reciprocal space via calculation of its diffraction pattern.
The latter is important because approximants can exhibit significant phason fluctuations (see discussion below).
We manually compare snapshots to tiling structures with the following wallpaper groups: p3m1 ($2/1$ approximant), p6m (other hexagonal $q/p$ approximants), p12m (dodecagonal QC), p4m (square lattice), and $E(2)$ (liquid).
Manual classification is typically unique and unambiguous.
In the presence of multiple tiling structures or in case defects remain in simulation snapshots, a snapshots is classified as belonging to the dominant tiling structure.

\begin{figure}
    \centering
    \includegraphics[width=0.8\linewidth]{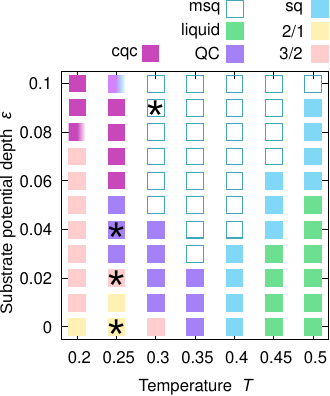}
    \caption{Assembly diagram of the LJG potential in the $T$-$\varepsilon$ parameter plane.
    Each data point corresponds to the dominant stucture observed in the final simulation snapshot after long relaxation.
    The ordered structures are: modulated square lattice (msq), square lattice (sq), $2/1$ approximant ($2/1$), $3/2$ approximant ($3/2$), dodecagonal QC (QC), and chiral structure (cqc).
    Snapshots at the parameter position indicated by four asterisks are shown in Fig.~\ref{fig:snapshots_examples_stabilityDiagram} and Fig.~\ref{fig:square_modulation}.}
    \label{fig:papermediumpotsimplified}
\end{figure}

\begin{figure*}
    \centering    
    \includegraphics[width=0.8\linewidth]{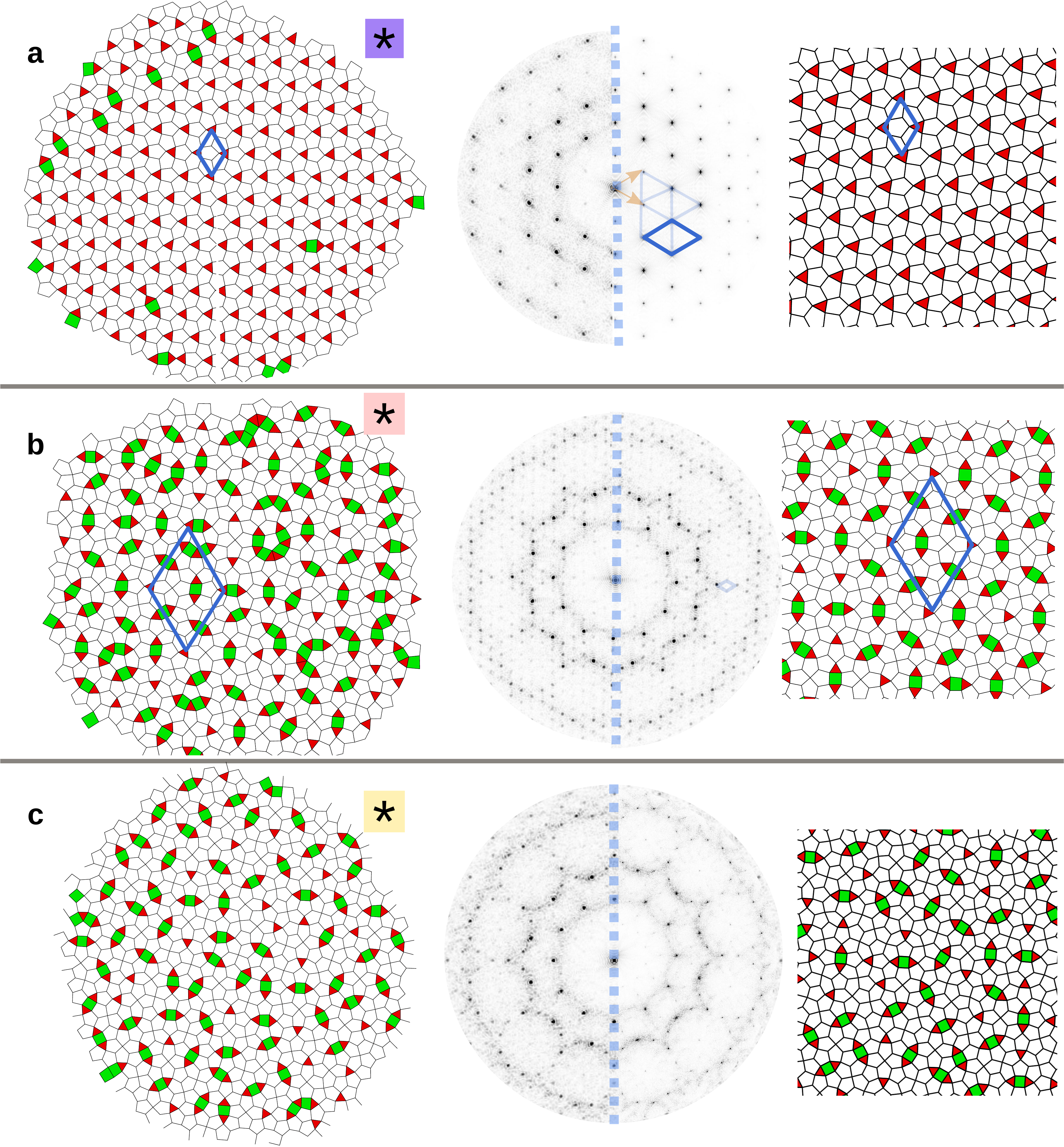}
    \caption{Snapshots from MD simulations at $T=0.25$ and at three substrate potential depths $\varepsilon$.
    Nearest neighbor bonds are shown only, with the particles omitted.
    Triangle tiles are shown in red, square tiles in green.
    \textbf{a}, $2/1$ approximant at $\varepsilon=0.00$.
    \textbf{b}, $3/2$ approximant at $\varepsilon=0.02$.
    \textbf{c}, QC at $\varepsilon=0.04$.
    Snapshots are shown in direct space obtained from simulation (left column) and from representative ideal tilings (right column).
    Diffraction patterns (middle column) are calculated and compared for both snapshots (left and right halves).
    The reciprocal lattice of diffraction spots in reciprocal space is mapped onto characteristic lattice vectors in direct space as indicated by blue lines in a, b.
    Notice the presence of significant phason mode fluctuations in the simulation snapshot of the $3/2$ approximant (b, left).
    Despite these phason mode fluctuations, its diffraction image (b, left half of middle) is highly similar to the diffraction image of the ideal tiling of the $3/2$ approximant without any phason mode fluctuations (b, right half of middle).}
    \label{fig:snapshots_examples_stabilityDiagram}
\end{figure*}

We summarize the structure classification in the assembly diagram of Fig.~\ref{fig:papermediumpotsimplified}.
In the absence of a substrate, $\varepsilon=0$, the known phase behavior of Eq.~(\ref{eq1}) is reproduced.
Notice that due to low resolution along the temperature axis, the continuous series of $q/p$ approximants reported previously~\cite{Engel_F-L_validation} is not fully resolved.
Phase boundaries shift with $\varepsilon$ according to four trends:
First, the dodecagonal QC and the $3/2$ approximant become stable over broader temperature ranges and shift towards lower temperature.
This finding demonstrates that a hexagonal substrate can, in fact, stabilize a dodecagonal quasicrystal despite it having lower symmetry than the quasicrystal.
Second, the stability region of the square lattice shifts towards high $T$ with increase of $\varepsilon$, and a new structure, called the modulated square lattice (msq), appears.
Third, melting temperature increases with $\varepsilon$.
Fourth, at low $T$ and high $\varepsilon$ we find a novel chiral quasicrystal (cqc).
In cqc, the inversion symmetry of the dodecgonal quasicrystal is broken.
Because this manuscript focuses on weak substrate potentials, cqc will be subject of a follow-up work.
Three simulation snapshots at parameters marked by asterisks in Fig.~\ref{fig:papermediumpotsimplified}, the $2/1$ approximant at $(T,\varepsilon)=(0.25,0.00)$, the $3/2$ approximant at $(0.25,0.02)$, and the quasicrystal at $(0.25,0.04)$ are shown in Fig.~\ref{fig:snapshots_examples_stabilityDiagram}.
A fourth simulation snapshot at parameters marked by an asterisk at $(0.3,0.09)$ corresponding to msq is shown in Fig.~\ref{fig:square_modulation}.

We make two further observations:
First, in the absence of a substrate, the phase transformation from the $2/1$ approximant to the QC was continuous via a series of $q/p$ approximants.
This continuous behavior is not sustained towards finite $\varepsilon$.
Instead, higher-order approximants are generally suppressed.
Specifically, we identify the low-order approximants $2/1$ and $3/2$ repeatedly in our snapshots for $\varepsilon>0$ but no other approximants.
Second, there remain significant phason mode fluctuations in the $3/2$ approximant at elevated temperatures.
Such phason mode fluctuations are apparent when comparing Fig.~\ref{fig:snapshots_examples_stabilityDiagram}b-left, which shows a simulation snapshot, to Fig.~\ref{fig:snapshots_examples_stabilityDiagram}b-right, which shows the ideal tiling of the $3/2$ approximant obtained after slow relaxation to $T=0$ to remove phason fluctuations.
Notice the similarity of the diffraction patterns calculated for both snapshots (two halves in Fig.~\ref{fig:snapshots_examples_stabilityDiagram}b-middle).
Apparently, the $3/2$ approximant behaves like a random tiling in that it permits phason mode fluctuations in thermodynamic equilibrium.
This means, we must rely on characterization of average crystallographic order by identification of diffraction spots in the diffraction pattern to distinguish the approximants in the generation of the assembly diagram.

The presence of phason mode fluctuations in an approximant is unexpected and to the best of our knowledge has not been reported.
It is unclear whether such behavior is related to the dimensionality of our system and can also occur in three dimensions.
We observe random tiling behavior only in sufficiently complex approximants.
For example, in the $2/1$ approximant there are no square tiles (green squares in Fig.~\ref{fig:snapshots_examples_stabilityDiagram}) except if connected to structural defects.
This means phason flips cannot occur inside a $2/1$ approximant but must be initiated from the surface or from a grain boundary.
In contrast, the $3/2$ approximant has many square tiles and thus readily supports phason mode fluctuations.

\subsection{Incommensurate-commensurate transition}

While the unperturbed (i.e., in the absence of a substrate) $2/1$ approximant transforms into the dodecagonal QC continuously via a series of hexagonal approximants, all approximants except $2/1$ and $3/2$ quickly disappear when the substrate is turned on (Fig.~\ref{fig:papermediumpotsimplified}).
Why is this the case?
To understand the disappearance of higher-order approximants, we resort to free energy calculations.
We calculate free energies for a well-relaxed QC and a number of low-order approximants.
Recall that hexagonal approximants are labeled by two integers as $q/p$ with $(\sqrt{3} + 1) / 2\approx1.37 < q/p < 2$.
Without loss of generality, we can require $q$ and $p$ to be coprime.
Finally, $q$ and $p$ should be small such that the approximants have small unit cells.
The simplest such approximants are $2/1 = 2$, $9/5 = 1.8$, $5/3 \approx 1.67$, $3/2 = 1.5$, and $7/5 = 1.4$.
We arrange the approximants in increasing order of $q/p$ ratio from the $2/1$ approximant, which is the energetic ground state, to the QC, which the equilibrium state at intermediate temperature.

\begin{figure*}
  \centering    
	\includegraphics[width=1\linewidth]{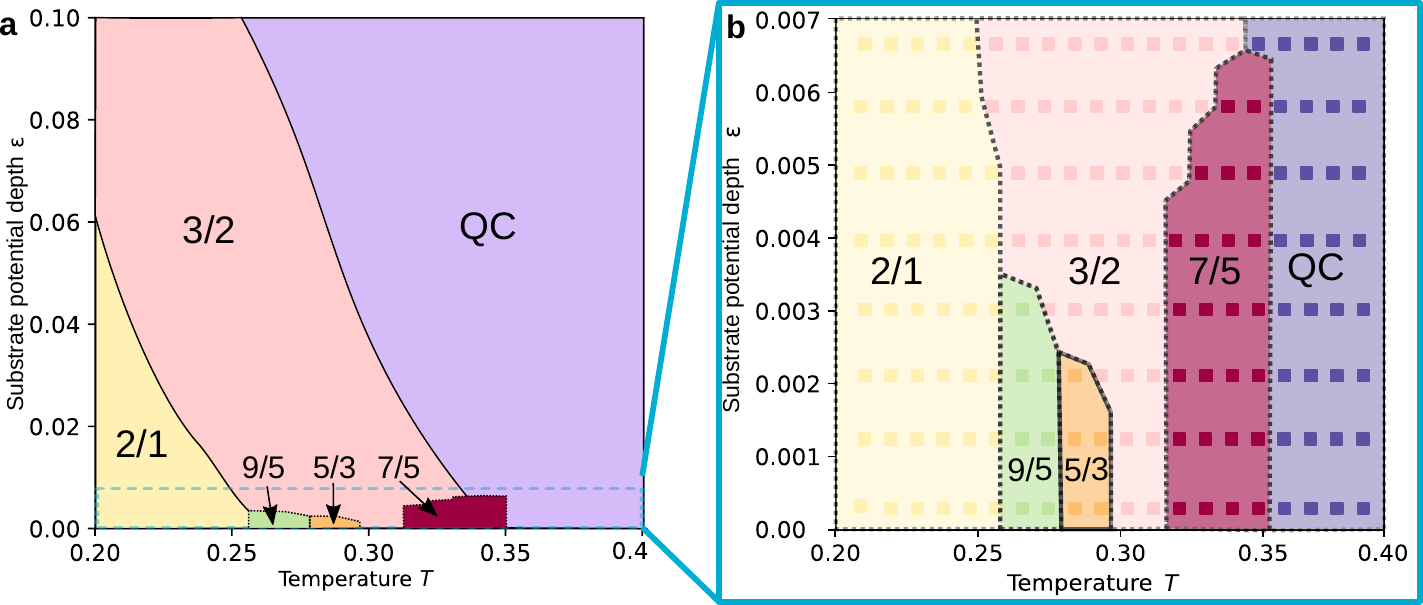}
	\caption{\textbf{a},~Phase diagram of the LJG potential with substrate from free energy calculations of the approximants $2/1$, $9/5$, $5/3$, $3/2$, $7/5$, and the QC as candidates.
    \textbf{b},~Zoom in on the region with weak substrate potential.
    The continuous transformation of the $2/1$ approximant into the QC transitions disappears as $\varepsilon$ increases.
    This phenomenon resembles the Devil's flower phenomenon known in modulated crystals where spontaneous phase-locking takes place.}
	\label{fig:comparison_of_periods}
\end{figure*}

Fig.~\ref{fig:comparison_of_periods} depicts the phase diagram predicted by free energy calculations.
In the absence of a substrate, $\varepsilon = 0$, the phase diagram cycles through the approximants $2/1$, $9/5$, $5/3$, $3/2$, $7/5$, and QC in correct (that is, in $q/p$ decreasing) order with temperature.
We also confirm the disappearance of all approximants except $2/1$ and $3/2$ with increasing $\varepsilon$.
Note that we did not include in the free energy calculations the square lattice and its modulation and the liquid.
Besides this omission, there is good agreement of the phase diagram in Fig.~\ref{fig:comparison_of_periods}
with the assembly diagram obtained from MD in Fig.~\ref{fig:papermediumpotsimplified}.

Situations where the competition of two length scales dominates phase behavior have long been studied in the field of modulated crystals\cite{Sander2007}.
Because the two length scales are of different physical or chemical origin, their ratio is not expected to be any special number and typically will be irrational.
The requirement of the system to satisfy both length scales simultaneously then causes the emergence of structural complexity in form of modulated crystals.
One distinguishes two cases:
Incommensurately modulated crystals are aperiodic crystals where both length scales are realized without compromise.
The dodecagonal QC in our system can be interpreted as an incommensurately modulated crystal.
A related example are Moiré patterns, where not a competition of length scales but the competition of crystallographic orientations and translation generates interference effects.
Incommensurately modulated crystals are accompanied by phason modes in thermodynamic equilibrium.
In contrast, commensurately modulated crystals are periodic crystals with large unit cells.
The lattice constants of commensurately modulated crystals simultaneously approximates integer multiples of both length scales.
Commensurately modulated crystals result from a lock-in transition where the irrational ratio of the two length scales locks into a rational approximation.
$q/p$ approximants are examples of commensurately modulated crystals.

Phase diagrams that include modulated crystals are usually studied as a function of length scale ratio and coupling strength.
The first parameter, length scale ratio, is often controlled indirectly by changing the composition or local structure.
In our system, length scale ratio is linked to temperature $T$ because temperature affects tile energy and thus the tile type that is dominant.
The second parameter, coupling strength, is measured relative to thermal energy.
In our system, coupling strength is given by substrate potential depth $\varepsilon$.
With these identifications, we can compare our phase diagram to past phase diagrams of modulated crystals.
The comparison shows that Fig.~\ref{fig:comparison_of_periods} closely resembles the mean field phase diagram for the 3D Ising model with competing interactions (Figure 24 in Ref.~\cite{PBak_1982}) and the FVdM model as suggested by Aubry (Figure 25 in Ref.~\cite{PBak_1982}).
Such characteristic phase diagrams of incommensurate-commensurate transitions have been called the \emph{Devil’s flower} phenomenon by Bak~\cite{PBak_1986}. Our results confirm that the increase of substrate strength locks modulated crystals into certain approximants and drives new complex phase behavior as expected from the Devil's flower phenomenon.

The appearance of the Devil's flower in our phase diagram helps us interpret the observed stabilization mechanism and particle dynamics.
In the absence of a substrate in the temperature range $0.24<T<0.36$, the system is incommensurately modulated.
Phason modes are unlocked and critical for thermodynamic stabilization because they contribute to free energy.
As substrate potential depth increases, phason modes may still be present (see Fig.~\ref{fig:snapshots_examples_stabilityDiagram}) but gradually disappear.
Eventually the system locks into the $2/1$ and $3/2$ approximants.
To the best of our knowledge, Fig.~\ref{fig:comparison_of_periods} is the first report of an incommensurate-commensurate transition in any two-dimensional system of freely moving particles.

\subsection{Square lattice and its variants}

\begin{figure*}
    \centering
    \includegraphics[width=1\linewidth]{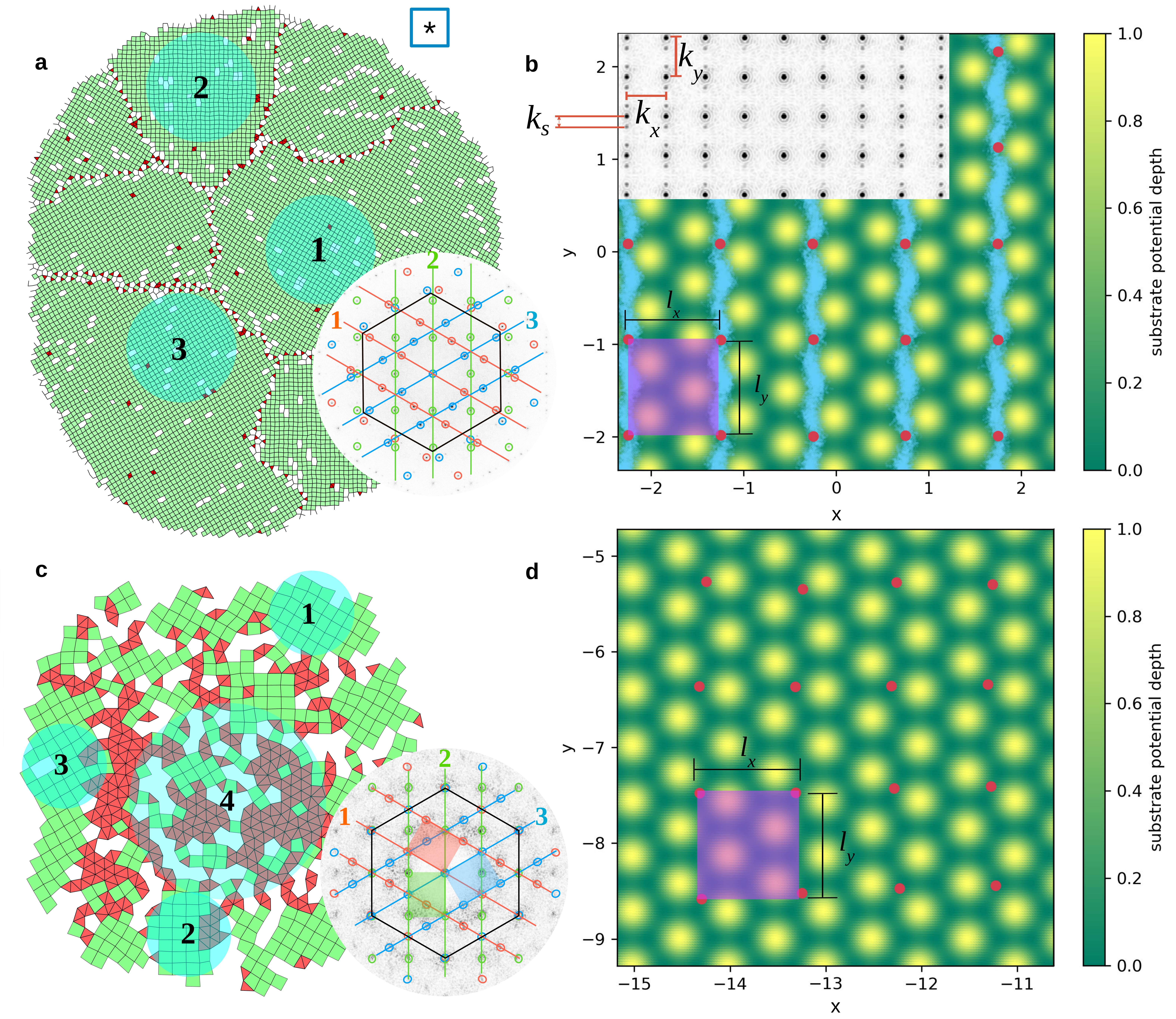}
    \caption{Variations of the square lattice at higher substrate potential depth $\varepsilon$ and intermediate temperature $T$.
    \textbf{a,b}, Modulated square lattice (msq) at $\varepsilon=0.09$ and $T=0.3$ for $10000$ particles.
    \textbf{c,d}, Coexistence of small grains with rectangular lattice (rec) and hexagonal lattice at $\varepsilon=1.0$ and $T=0.4$ for $1024$ particles.
    \textbf{a,c}, Particle configurations showing nearest neighbor bonds only, with the particles omitted.
    Triangle tiles are shown in red, square tiles in green.
    There coexist grains in three orientations (labelled '1', '2', '3') and a region dominated by triangles (labelled '4' in (d)).
    Diffraction pattern are shown as bottom right insets with diffraction peaks encircled and colored according to the grain orientation they correspond to.
    \textbf{b,d}, Zoom ins on the grain '2'. Particles are shown as red disks overlaid on the substrate potential.
    In (b), light blue color indicates the channels the particles move in.
    A diffraction pattern is shown as top left inset.
    }
	\label{fig:square_modulation}
\end{figure*}

The presence of the substrate drives the incommensurate-commensurate transition at low substrate potential depth $\varepsilon<0.03$.
For $\varepsilon>0.03$, the square lattice (sq) and its variants appear dominantly in the phase diagram (Fig.~\ref{fig:papermediumpotsimplified}).
For intermediate substrate potentials, a novel modulated square lattice (msq) is observed (Fig.~\ref{fig:square_modulation}a,b).
For high substrate potential, $\varepsilon\gg1.0$, above the range covered in Fig.~\ref{fig:papermediumpotsimplified}, a coexistence of grains with rectangular lattice (rec) and hexagonal lattice is eventually observed (Fig.~\ref{fig:square_modulation}c,d).
All of these phenomena are yet other examples of commensurate-incommensurate transitions induced by the competition of the length scales imparted by the LJG interaction potential with the unit cell of the substrate.

Msq and rec are typically polycrystalline with individual grains locking into one of three distinct orientations.
This locking is a consequence of the six-fold symmetry of the substrate.
Diffraction patterns confirm the six-fold symmetry.
In the diffraction pattern of msq (inset of Fig.~\ref{fig:square_modulation}a), isolated peaks (circled) are present for each grain.
Besides the locking of grain orientations, individual grains are only weakly affected by the substrate (see modulations discussed below).
In the diffraction pattern of rec (inset of 
Fig.~\ref{fig:square_modulation}c), some peak triplets merge into single peaks (indicated by black hexagons in Fig.~\ref{fig:square_modulation}a,c).
The merging is caused by a stretching of squares into rectangles to improve compatibility with the substrate.
This means four-fold symmetry is already lost in individual grain of rec and the system is closer to six-fold symmetry than msq.
While small patches of triangles are sometimes already present in rec (central part of Fig.~\ref{fig:square_modulation}c labelled '4'), only very high $\varepsilon$ transforms the system completely into a hexagonal crystal (not shown).

A zoom-in on grain '2' of rec is shown in Fig.~\ref{fig:square_modulation}d.
We plot the substrate potential in the background to visualize particle locations (red) relative to maxima (yellow) and minima (green) of the potential.
Potential maxima form a hexagonal lattice separated by channels of low potential that trace the nearest-neighbor bond network of a honeycomb lattice.
The stretching of squares into rectangles can now be understood.
Recall the nearest-neighbor distance is $r_\text{NN}\approx1.0$, and the substrate periodicity is chosen as $\lambda=0.5$.
A lattice constant $l_x=2\lambda$ allows particles to retain their preferred nearest-neighbor distance without penalty.
And a slight adjustment of the lattice constant to $l_y=\sqrt{4/3} l_x\approx 1.15 l_x$ ensures commensurability with the substrate also in the $y$-direction.

A zoom-in on grain '2' of msq is shown in Fig.~\ref{fig:square_modulation}b.
Just like before, a lattice constant $l_x=2\lambda$ allows particles to retain their preferred nearest-neighbor distance without penalty.
The $y$-direction is again more complicated.
A lattice constant $l_y=l_x$ now guarantees four-fold symmetry in the diffraction pattern but can do so only in average and by allowing positional modulations along the $y$-axis.
A close inspection reveals the presence of these modulation in form of weak satellite peaks in the diffraction pattern (inset of Fig.~\ref{fig:square_modulation}b).
The separation of satellite peaks to the central peaks corresponds to a superstructure periodicity of $l_s\approx0.20$.
Analysis of other snapshots suggests that $l_s$ is not unique but a function of $T$ and $\varepsilon$.
We also observe that particles are unusually mobile in channels along the $y$-direction (purple in Fig.~\ref{fig:square_modulation}b).
This high mobility can be directly associated with a phason mode.
All evidence combined demonstrates that msq is an incommensurately modulated crystal.

\section{Conclusions}

Research in this paper was originally motivated by the discovery of approximants in the thin-film perovskites (Ba, Sc)TiO$_{3}$ on solid metal substrates.
Importantly, experiments demonstrated that the type of approximant changes as a function of substrate interaction and periodicity.
Our findings reproduce and explain aspects of these experimental observations.
Our findings also highlight a number of new phenomena, not yet seen in experiment or simulation.

The new phenomena we observed, and that we believe could be general consequences of the competition between a QC and a periodic substrate, are:
\begin{enumerate}
\item Existence of incommensurate-commensurate phase transitions. In these phase transitions, the QC locks into one or several approximants.
We observed the stabilization of the $3/2$ approximant via a lock-in transition.
\item Approximants with phason fluctuations.
We observed that the $3/2$ approximant resembles a random tiling rather than an ideal crystal despite having a well-defined periodicity as indicated by the presence of a reciprocal lattice.
\item Enhanced stability of the QC.
Counter-intuitively, the substrate enhanced the thermodynamic stability of the dodecagonal quasicrystal for some parameter values.
\item Incommensurately modulated crystals, where the modulation is a consequence of the competition between two length scales.
We observed the modulated square lattice.
\item Novel structures commensurate with the substrate that are not approximants.
We observed the rectangular lattice.
\end{enumerate}
While observation 1 is similar to that in thin-film perovskites and observation 3 could explain the existence of perovskite QCs, the other observations have not yet been reported in the experimental system.
It will be interesting to search for them.

Generally, we see a trend that the competition between the inherent symmetry of the model system, i.e., the crystal structure the system prefers to form in the absence of a substrate, and the imposed symmetry of the substrate can occur in multiple steps involving both commensurate and incommensurate phases.
We find in our system, as we increase the strength of the substrate potential, at $T=0.25$ (see Fig.~\ref{fig:papermediumpotsimplified}) a complex sequence of phase transformation alternating multiple times between commensurate and incommensurate: from (i)~2/1 approximant (crystal) to (ii)~3/2 approximant (commensurate) to (iii)~quasicrystal (incommensurate) to (iv)~modulated square lattice (incommensurate) to (v)~rectangle lattice (commensurate) to (vi)~hexagonal lattice (substrate).

Future work should further expand this research by testing more parameters or conducting similar studies in more realistic systems.
An important parameter that should be varied, but was not in this work, is the substrate periodicity.
Varying this parameter, it should be possible to target desired crystal structures and approximants by inverse design.
Given that many recent model systems of quasicrystals have been found in soft matter systems and the fact that such systems are highly tunable and thus ideally suited for parameter studies, we believe soft matter quasicrystals are ideal candidates to test many of the discoveries reported in this work.

\section*{Author contributions}
N.R.V.-R. implemented the model and performed all simulations and free energy calculations.
Both authors conceived the idea, discussed the results, and wrote the manuscript.
M.E. developed code to construct tilings and supervised the project.

\section*{Conflicts of interest}
There are no conflicts to declare.

\section*{Acknowledgements}
We acknowledge funding by the Deutsche Forschungsgemeinschaft (DFG) through grant EN 905/4-1. The authors gratefully acknowledge the HPC resources provided by the Erlangen National High Performance Computing Center (NHR@FAU) of the Friedrich-Alexander-Universität Erlangen-Nürnberg (FAU) under the NHR project CRC1411D04.

\balance
\bibliography{QuasicrystalOnSubstrate_Manuscript_revision2} 

\providecommand*{\mcitethebibliography}{\thebibliography}
\csname @ifundefined\endcsname{endmcitethebibliography}
{\let\endmcitethebibliography\endthebibliography}{}
\begin{mcitethebibliography}{54}
\providecommand*{\natexlab}[1]{#1}
\providecommand*{\mciteSetBstSublistMode}[1]{}
\providecommand*{\mciteSetBstMaxWidthForm}[2]{}
\providecommand*{\mciteBstWouldAddEndPuncttrue}
  {\def\EndOfBibitem{\unskip.}}
\providecommand*{\mciteBstWouldAddEndPunctfalse}
  {\let\EndOfBibitem\relax}
\providecommand*{\mciteSetBstMidEndSepPunct}[3]{}
\providecommand*{\mciteSetBstSublistLabelBeginEnd}[3]{}
\providecommand*{\EndOfBibitem}{}
\mciteSetBstSublistMode{f}
\mciteSetBstMaxWidthForm{subitem}
{(\emph{\alph{mcitesubitemcount}})}
\mciteSetBstSublistLabelBeginEnd{\mcitemaxwidthsubitemform\space}
{\relax}{\relax}

\bibitem[Shechtman \emph{et~al.}(1984)Shechtman, Blech, Gratias, and
  Cahn]{Shechtman_1983}
D.~Shechtman, I.~Blech, D.~Gratias and J.~W. Cahn, \emph{Phys. Rev. Lett.},
  1984, \textbf{53}, 1951--1953\relax
\mciteBstWouldAddEndPuncttrue
\mciteSetBstMidEndSepPunct{\mcitedefaultmidpunct}
{\mcitedefaultendpunct}{\mcitedefaultseppunct}\relax
\EndOfBibitem
\bibitem[Levine and Steinhardt(1984)]{Levine_1984}
D.~Levine and P.~J. Steinhardt, \emph{Phys. Rev. Lett.}, 1984, \textbf{53},
  2477--2480\relax
\mciteBstWouldAddEndPuncttrue
\mciteSetBstMidEndSepPunct{\mcitedefaultmidpunct}
{\mcitedefaultendpunct}{\mcitedefaultseppunct}\relax
\EndOfBibitem
\bibitem[Steurer(2018)]{Steurer_2018}
W.~Steurer, \emph{Acta Crystallogr., Sect. A}, 2018, \textbf{74}, 1--11\relax
\mciteBstWouldAddEndPuncttrue
\mciteSetBstMidEndSepPunct{\mcitedefaultmidpunct}
{\mcitedefaultendpunct}{\mcitedefaultseppunct}\relax
\EndOfBibitem
\bibitem[Janot(1996)]{thermalConductivity_1996}
C.~Janot, \emph{Phys. Rev. B}, 1996, \textbf{53}, 181--191\relax
\mciteBstWouldAddEndPuncttrue
\mciteSetBstMidEndSepPunct{\mcitedefaultmidpunct}
{\mcitedefaultendpunct}{\mcitedefaultseppunct}\relax
\EndOfBibitem
\bibitem[Takeuchi \emph{et~al.}(1991)Takeuchi, Iwanaga, and
  Shibuya]{coating_pt2_1991}
S.~Takeuchi, H.~Iwanaga and T.~Shibuya, \emph{Jpn. J. Appl. Phys.}, 1991,
  \textbf{30}, 561\relax
\mciteBstWouldAddEndPuncttrue
\mciteSetBstMidEndSepPunct{\mcitedefaultmidpunct}
{\mcitedefaultendpunct}{\mcitedefaultseppunct}\relax
\EndOfBibitem
\bibitem[Sha{\u{\i}}tura and Enaleeva(2007)]{coating_2007}
D.~S. Sha{\u{\i}}tura and A.~A. Enaleeva, \emph{Crystallogr. Rep.}, 2007,
  \textbf{52}, 945--952\relax
\mciteBstWouldAddEndPuncttrue
\mciteSetBstMidEndSepPunct{\mcitedefaultmidpunct}
{\mcitedefaultendpunct}{\mcitedefaultseppunct}\relax
\EndOfBibitem
\bibitem[Vardeny \emph{et~al.}(2013)Vardeny, Nahata, and
  Agrawal]{photonic_Vardeny2013}
Z.~V. Vardeny, A.~Nahata and A.~Agrawal, \emph{Nat. Photonics}, 2013,
  \textbf{7}, 177--187\relax
\mciteBstWouldAddEndPuncttrue
\mciteSetBstMidEndSepPunct{\mcitedefaultmidpunct}
{\mcitedefaultendpunct}{\mcitedefaultseppunct}\relax
\EndOfBibitem
\bibitem[Kamiya \emph{et~al.}(2018)Kamiya, Takeuchi, Kabeya, Wada, Ishimasa,
  Ochiai, Deguchi, Imura, and Sato]{SuperCond_Kamiya2018}
K.~Kamiya, T.~Takeuchi, N.~Kabeya, N.~Wada, T.~Ishimasa, A.~Ochiai, K.~Deguchi,
  K.~Imura and N.~K. Sato, \emph{Nat. Commun.}, 2018, \textbf{9}, 154\relax
\mciteBstWouldAddEndPuncttrue
\mciteSetBstMidEndSepPunct{\mcitedefaultmidpunct}
{\mcitedefaultendpunct}{\mcitedefaultseppunct}\relax
\EndOfBibitem
\bibitem[F{\"o}rster \emph{et~al.}(2013)F{\"o}rster, Meinel, Hammer, Trautmann,
  and Widdra]{Nature_Forster2013}
S.~F{\"o}rster, K.~Meinel, R.~Hammer, M.~Trautmann and W.~Widdra,
  \emph{Nature}, 2013, \textbf{502}, 215--218\relax
\mciteBstWouldAddEndPuncttrue
\mciteSetBstMidEndSepPunct{\mcitedefaultmidpunct}
{\mcitedefaultendpunct}{\mcitedefaultseppunct}\relax
\EndOfBibitem
\bibitem[F\"orster \emph{et~al.}(2016)F\"orster, Trautmann, Roy, Adeagbo,
  Zollner, Hammer, Schumann, Meinel, Nayak, Mohseni, Hergert, Meyerheim, and
  Widdra]{Foerster_OQC_2016}
S.~F\"orster, M.~Trautmann, S.~Roy, W.~A. Adeagbo, E.~M. Zollner, R.~Hammer,
  F.~O. Schumann, K.~Meinel, S.~K. Nayak, K.~Mohseni, W.~Hergert, H.~L.
  Meyerheim and W.~Widdra, \emph{Phys. Rev. Lett.}, 2016, \textbf{117},
  095501\relax
\mciteBstWouldAddEndPuncttrue
\mciteSetBstMidEndSepPunct{\mcitedefaultmidpunct}
{\mcitedefaultendpunct}{\mcitedefaultseppunct}\relax
\EndOfBibitem
\bibitem[Foerster \emph{et~al.}(2017)Foerster, Flege, Falta, Zollner, Schumann,
  Hammer, Bayat, Schindler, and Widdra]{Foerster2017}
S.~Foerster, J.~I. Flege, J.~Falta, E.~M. Zollner, F.~O. Schumann, R.~Hammer,
  A.~Bayat, K.-M. Schindler and W.~Widdra, \emph{Ann. Phys.}, 2017,
  \textbf{529}, 1--7\relax
\mciteBstWouldAddEndPuncttrue
\mciteSetBstMidEndSepPunct{\mcitedefaultmidpunct}
{\mcitedefaultendpunct}{\mcitedefaultseppunct}\relax
\EndOfBibitem
\bibitem[Förster \emph{et~al.}(2020)Förster, Schenk, Maria~Zollner, Krahn,
  Chiang, Schumann, Bayat, Schindler, Trautmann, Hammer, Meinel, Adeagbo,
  Hergert, Ingo~Flege, Falta, Ellguth, Tusche, DeBoissieu, Muntwiler, Greber,
  and Widdra]{Foerster_2019}
S.~Förster, S.~Schenk, E.~Maria~Zollner, O.~Krahn, C.-T. Chiang, F.~O.
  Schumann, A.~Bayat, K.-M. Schindler, M.~Trautmann, R.~Hammer, K.~Meinel,
  W.~A. Adeagbo, W.~Hergert, J.~Ingo~Flege, J.~Falta, M.~Ellguth, C.~Tusche,
  M.~DeBoissieu, M.~Muntwiler, T.~Greber and W.~Widdra, \emph{Phys. Status
  Solidi B}, 2020, \textbf{257}, 1900624\relax
\mciteBstWouldAddEndPuncttrue
\mciteSetBstMidEndSepPunct{\mcitedefaultmidpunct}
{\mcitedefaultendpunct}{\mcitedefaultseppunct}\relax
\EndOfBibitem
\bibitem[Maniraj \emph{et~al.}(2021)Maniraj, Tran, Krahn, Schenk, Widdra, and
  F\"orster]{Foerster_2021}
M.~Maniraj, L.~V. Tran, O.~Krahn, S.~Schenk, W.~Widdra and S.~F\"orster,
  \emph{Phys. Rev. Mater.}, 2021, \textbf{5}, 084006\relax
\mciteBstWouldAddEndPuncttrue
\mciteSetBstMidEndSepPunct{\mcitedefaultmidpunct}
{\mcitedefaultendpunct}{\mcitedefaultseppunct}\relax
\EndOfBibitem
\bibitem[Zollner \emph{et~al.}(2020)Zollner, Schenk, Setvin, and
  Förster]{Zollner_2020}
E.~M. Zollner, S.~Schenk, M.~Setvin and S.~Förster, \emph{Phys. Status Solidi
  B}, 2020, \textbf{257}, 1900620\relax
\mciteBstWouldAddEndPuncttrue
\mciteSetBstMidEndSepPunct{\mcitedefaultmidpunct}
{\mcitedefaultendpunct}{\mcitedefaultseppunct}\relax
\EndOfBibitem
\bibitem[Maria~Zollner \emph{et~al.}(2020)Maria~Zollner, Schuster, Meinel,
  Stötzner, Schenk, Allner, Förster, and Widdra]{Zollner_2020_pt2}
E.~Maria~Zollner, F.~Schuster, K.~Meinel, P.~Stötzner, S.~Schenk, B.~Allner,
  S.~Förster and W.~Widdra, \emph{Phys. Status Solidi B}, 2020, \textbf{257},
  1900655\relax
\mciteBstWouldAddEndPuncttrue
\mciteSetBstMidEndSepPunct{\mcitedefaultmidpunct}
{\mcitedefaultendpunct}{\mcitedefaultseppunct}\relax
\EndOfBibitem
\bibitem[Cockayne \emph{et~al.}(2016)Cockayne, Mihalkovi\ifmmode~\check{c}\else
  \v{c}\fi{}, and Henley]{Mihalkovik_2016}
E.~Cockayne, M.~Mihalkovi\ifmmode~\check{c}\else \v{c}\fi{} and C.~L. Henley,
  \emph{Phys. Rev. B}, 2016, \textbf{93}, 020101\relax
\mciteBstWouldAddEndPuncttrue
\mciteSetBstMidEndSepPunct{\mcitedefaultmidpunct}
{\mcitedefaultendpunct}{\mcitedefaultseppunct}\relax
\EndOfBibitem
\bibitem[Dorini \emph{et~al.}(2021)Dorini, Brix, Chatelier, Kokalj, and
  Gaudry]{Thiago_2021}
T.~T. Dorini, F.~Brix, C.~Chatelier, A.~Kokalj and E.~Gaudry, \emph{Nanoscale},
  2021, \textbf{13}, 10771--10779\relax
\mciteBstWouldAddEndPuncttrue
\mciteSetBstMidEndSepPunct{\mcitedefaultmidpunct}
{\mcitedefaultendpunct}{\mcitedefaultseppunct}\relax
\EndOfBibitem
\bibitem[Patrykiejew \emph{et~al.}(2009)Patrykiejew, R{\.{z}}ysko, and
  Soko{\l}owski]{monolayer_Patrykiejew2009}
A.~Patrykiejew, W.~R{\.{z}}ysko and S.~Soko{\l}owski, \emph{Adsorpt.}, 2009,
  \textbf{15}, 254--263\relax
\mciteBstWouldAddEndPuncttrue
\mciteSetBstMidEndSepPunct{\mcitedefaultmidpunct}
{\mcitedefaultendpunct}{\mcitedefaultseppunct}\relax
\EndOfBibitem
\bibitem[Neuhaus \emph{et~al.}(2013)Neuhaus, Marechal, Schmiedeberg, and
  L\"owen]{sq_sub_Schmiedeberg_2013}
T.~Neuhaus, M.~Marechal, M.~Schmiedeberg and H.~L\"owen, \emph{Phys. Rev.
  Lett.}, 2013, \textbf{110}, 118301\relax
\mciteBstWouldAddEndPuncttrue
\mciteSetBstMidEndSepPunct{\mcitedefaultmidpunct}
{\mcitedefaultendpunct}{\mcitedefaultseppunct}\relax
\EndOfBibitem
\bibitem[Brazda \emph{et~al.}(2018)Brazda, Silva, Manini, Vanossi, Guerra,
  Tosatti, and Bechinger]{experiment_2dAburby_2018}
T.~Brazda, A.~Silva, N.~Manini, A.~Vanossi, R.~Guerra, E.~Tosatti and
  C.~Bechinger, \emph{Phys. Rev. X}, 2018, \textbf{8}, 011050\relax
\mciteBstWouldAddEndPuncttrue
\mciteSetBstMidEndSepPunct{\mcitedefaultmidpunct}
{\mcitedefaultendpunct}{\mcitedefaultseppunct}\relax
\EndOfBibitem
\bibitem[Baykara \emph{et~al.}(2018)Baykara, Vazirisereshk, and
  Martini]{superlubricity_review_2018}
M.~Z. Baykara, M.~R. Vazirisereshk and A.~Martini, \emph{Appl. Phys. Rev.},
  2018, \textbf{5}, 041102\relax
\mciteBstWouldAddEndPuncttrue
\mciteSetBstMidEndSepPunct{\mcitedefaultmidpunct}
{\mcitedefaultendpunct}{\mcitedefaultseppunct}\relax
\EndOfBibitem
\bibitem[Amiri \emph{et~al.}(2021)Amiri, Volkert, and
  Vink]{Vink_incommensurateSubstrate}
S.~Amiri, C.~A. Volkert and R.~L.~C. Vink, \emph{Phys. Rev. E}, 2021,
  \textbf{104}, 014802\relax
\mciteBstWouldAddEndPuncttrue
\mciteSetBstMidEndSepPunct{\mcitedefaultmidpunct}
{\mcitedefaultendpunct}{\mcitedefaultseppunct}\relax
\EndOfBibitem
\bibitem[Cao \emph{et~al.}(2021)Cao, Panizon, Vanossi, Manini, Tosatti, and
  Bechinger]{Sliding_substrate_2021}
X.~Cao, E.~Panizon, A.~Vanossi, N.~Manini, E.~Tosatti and C.~Bechinger,
  \emph{Phys. Rev. E}, 2021, \textbf{103}, 012606\relax
\mciteBstWouldAddEndPuncttrue
\mciteSetBstMidEndSepPunct{\mcitedefaultmidpunct}
{\mcitedefaultendpunct}{\mcitedefaultseppunct}\relax
\EndOfBibitem
\bibitem[Reichhardt and Reichhardt(2016)]{review_2016_Reichhardt_depinning}
C.~Reichhardt and C.~J.~O. Reichhardt, \emph{Rep. Prog. Phys.}, 2016,
  \textbf{80}, 026501\relax
\mciteBstWouldAddEndPuncttrue
\mciteSetBstMidEndSepPunct{\mcitedefaultmidpunct}
{\mcitedefaultendpunct}{\mcitedefaultseppunct}\relax
\EndOfBibitem
\bibitem[Flückiger \emph{et~al.}(2003)Flückiger, Weisskopf, Erbudak,
  Lüscher, and Kortan]{experimental_ex_nanoepitaxy}
T.~Flückiger, Y.~Weisskopf, M.~Erbudak, R.~Lüscher and A.~R. Kortan,
  \emph{Nano Lett.}, 2003, \textbf{3}, 1717--1721\relax
\mciteBstWouldAddEndPuncttrue
\mciteSetBstMidEndSepPunct{\mcitedefaultmidpunct}
{\mcitedefaultendpunct}{\mcitedefaultseppunct}\relax
\EndOfBibitem
\bibitem[Ferralis \emph{et~al.}(2004)Ferralis, Diehl, Pussi, Lindroos, Fisher,
  and Jenks]{experimental_ex_LEED}
N.~Ferralis, R.~D. Diehl, K.~Pussi, M.~Lindroos, I.~Fisher and C.~J. Jenks,
  \emph{Phys. Rev. B}, 2004, \textbf{69}, 075410\relax
\mciteBstWouldAddEndPuncttrue
\mciteSetBstMidEndSepPunct{\mcitedefaultmidpunct}
{\mcitedefaultendpunct}{\mcitedefaultseppunct}\relax
\EndOfBibitem
\bibitem[Bilki \emph{et~al.}(2007)Bilki, Erbudak, Mungan, and
  Weisskopf]{Weisskopf_Y_sim_sub_deca_2007}
B.~Bilki, M.~Erbudak, M.~Mungan and Y.~Weisskopf, \emph{Phys. Rev. B}, 2007,
  \textbf{75}, 045437\relax
\mciteBstWouldAddEndPuncttrue
\mciteSetBstMidEndSepPunct{\mcitedefaultmidpunct}
{\mcitedefaultendpunct}{\mcitedefaultseppunct}\relax
\EndOfBibitem
\bibitem[Smerdon \emph{et~al.}(2014)Smerdon, Young, Lowe, Hars, Yadav, Hesp,
  Dhanak, Tsai, Sharma, and McGrath]{McGrath_templated_QC_mol_ord_2014}
J.~A. Smerdon, K.~M. Young, M.~Lowe, S.~S. Hars, T.~P. Yadav, D.~Hesp, V.~R.
  Dhanak, A.~P. Tsai, H.~R. Sharma and R.~McGrath, \emph{Nano Lett.}, 2014,
  \textbf{14}, 1184--1189\relax
\mciteBstWouldAddEndPuncttrue
\mciteSetBstMidEndSepPunct{\mcitedefaultmidpunct}
{\mcitedefaultendpunct}{\mcitedefaultseppunct}\relax
\EndOfBibitem
\bibitem[Mikhael \emph{et~al.}(2008)Mikhael, Roth, Helden, and
  Bechinger]{arch_tiles_Mikhael2008}
J.~Mikhael, J.~Roth, L.~Helden and C.~Bechinger, \emph{Nature}, 2008,
  \textbf{454}, 501--504\relax
\mciteBstWouldAddEndPuncttrue
\mciteSetBstMidEndSepPunct{\mcitedefaultmidpunct}
{\mcitedefaultendpunct}{\mcitedefaultseppunct}\relax
\EndOfBibitem
\bibitem[Schmiedeberg and Stark(2008)]{Schmiedeberg_2D_quasicrystSubst}
M.~Schmiedeberg and H.~Stark, \emph{Phys. Rev. Lett.}, 2008, \textbf{101},
  218302\relax
\mciteBstWouldAddEndPuncttrue
\mciteSetBstMidEndSepPunct{\mcitedefaultmidpunct}
{\mcitedefaultendpunct}{\mcitedefaultseppunct}\relax
\EndOfBibitem
\bibitem[Schmiedeberg \emph{et~al.}(2010)Schmiedeberg, Mikhael, Rausch, Roth,
  Helden, Bechinger, and Stark]{arch_tiles_part2_Schmiedeberg2010}
M.~Schmiedeberg, J.~Mikhael, S.~Rausch, J.~Roth, L.~Helden, C.~Bechinger and
  H.~Stark, \emph{Eur. Phys. J. E}, 2010, \textbf{32}, 25--34\relax
\mciteBstWouldAddEndPuncttrue
\mciteSetBstMidEndSepPunct{\mcitedefaultmidpunct}
{\mcitedefaultendpunct}{\mcitedefaultseppunct}\relax
\EndOfBibitem
\bibitem[Mikhael \emph{et~al.}(2011)Mikhael, Gera, Bohlein, and
  Bechinger]{arch_tiles_Mikhael}
J.~Mikhael, G.~Gera, T.~Bohlein and C.~Bechinger, \emph{Soft Matter}, 2011,
  \textbf{7}, 1352--1357\relax
\mciteBstWouldAddEndPuncttrue
\mciteSetBstMidEndSepPunct{\mcitedefaultmidpunct}
{\mcitedefaultendpunct}{\mcitedefaultseppunct}\relax
\EndOfBibitem
\bibitem[Jagannathan and Duneau(2014)]{Jagannathan2014_eight_fold_laser}
A.~Jagannathan and M.~Duneau, \emph{Eur. Phys. J. B}, 2014, \textbf{87},
  149\relax
\mciteBstWouldAddEndPuncttrue
\mciteSetBstMidEndSepPunct{\mcitedefaultmidpunct}
{\mcitedefaultendpunct}{\mcitedefaultseppunct}\relax
\EndOfBibitem
\bibitem[Leung \emph{et~al.}(1989)Leung, Henley, and
  Chester]{Leung_dod-QC_1989}
P.~W. Leung, C.~L. Henley and G.~V. Chester, \emph{Phys. Rev. B}, 1989,
  \textbf{39}, 446--458\relax
\mciteBstWouldAddEndPuncttrue
\mciteSetBstMidEndSepPunct{\mcitedefaultmidpunct}
{\mcitedefaultendpunct}{\mcitedefaultseppunct}\relax
\EndOfBibitem
\bibitem[Fayen \emph{et~al.}(2023)Fayen, Impéror-Clerc, Filion, Foffi, and
  Smallenburg]{Fayen2023}
E.~Fayen, M.~Impéror-Clerc, L.~Filion, G.~Foffi and F.~Smallenburg, \emph{Soft
  Matter}, 2023, \textbf{19}, 2654--2663\relax
\mciteBstWouldAddEndPuncttrue
\mciteSetBstMidEndSepPunct{\mcitedefaultmidpunct}
{\mcitedefaultendpunct}{\mcitedefaultseppunct}\relax
\EndOfBibitem
\bibitem[Quandt and Teter(1999)]{Quandt_1999}
A.~Quandt and M.~P. Teter, \emph{Phys. Rev. B}, 1999, \textbf{59},
  8586--8592\relax
\mciteBstWouldAddEndPuncttrue
\mciteSetBstMidEndSepPunct{\mcitedefaultmidpunct}
{\mcitedefaultendpunct}{\mcitedefaultseppunct}\relax
\EndOfBibitem
\bibitem[Engel and Trebin(2007)]{Engel_2007}
M.~Engel and H.-R. Trebin, \emph{Phys. Rev. Lett.}, 2007, \textbf{98},
  225505\relax
\mciteBstWouldAddEndPuncttrue
\mciteSetBstMidEndSepPunct{\mcitedefaultmidpunct}
{\mcitedefaultendpunct}{\mcitedefaultseppunct}\relax
\EndOfBibitem
\bibitem[Engel \emph{et~al.}(2010)Engel, Umezaki, Trebin, and
  Odagaki]{Engel_2010}
M.~Engel, M.~Umezaki, H.-R. Trebin and T.~Odagaki, \emph{Phys. Rev. B}, 2010,
  \textbf{82}, 134206\relax
\mciteBstWouldAddEndPuncttrue
\mciteSetBstMidEndSepPunct{\mcitedefaultmidpunct}
{\mcitedefaultendpunct}{\mcitedefaultseppunct}\relax
\EndOfBibitem
\bibitem[van~der Linden \emph{et~al.}(2012)van~der Linden, Doye, and
  Louis]{Marjolein_2012}
M.~N. van~der Linden, J.~P.~K. Doye and A.~A. Louis, \emph{J. Chem. Phys.},
  2012, \textbf{136}, 054904\relax
\mciteBstWouldAddEndPuncttrue
\mciteSetBstMidEndSepPunct{\mcitedefaultmidpunct}
{\mcitedefaultendpunct}{\mcitedefaultseppunct}\relax
\EndOfBibitem
\bibitem[Reinhardt \emph{et~al.}(2013)Reinhardt, Romano, and Doye]{Doye2013}
A.~Reinhardt, F.~Romano and J.~P.~K. Doye, \emph{Phys. Rev. Lett.}, 2013,
  \textbf{110}, 255503\relax
\mciteBstWouldAddEndPuncttrue
\mciteSetBstMidEndSepPunct{\mcitedefaultmidpunct}
{\mcitedefaultendpunct}{\mcitedefaultseppunct}\relax
\EndOfBibitem
\bibitem[Engel(2011)]{Engel_F-L_validation}
M.~Engel, \emph{Phys. Rev. Lett.}, 2011, \textbf{106}, 095504\relax
\mciteBstWouldAddEndPuncttrue
\mciteSetBstMidEndSepPunct{\mcitedefaultmidpunct}
{\mcitedefaultendpunct}{\mcitedefaultseppunct}\relax
\EndOfBibitem
\bibitem[Oxborrow and Henley(1993)]{Henley1993}
M.~Oxborrow and C.~L. Henley, \emph{Phys. Rev. B}, 1993, \textbf{48},
  6966--6998\relax
\mciteBstWouldAddEndPuncttrue
\mciteSetBstMidEndSepPunct{\mcitedefaultmidpunct}
{\mcitedefaultendpunct}{\mcitedefaultseppunct}\relax
\EndOfBibitem
\bibitem[{de Bruijn}(1981)]{deBruijn1981}
N.~{de Bruijn}, \emph{Indagationes Math.}, 1981, \textbf{84}, 39--52\relax
\mciteBstWouldAddEndPuncttrue
\mciteSetBstMidEndSepPunct{\mcitedefaultmidpunct}
{\mcitedefaultendpunct}{\mcitedefaultseppunct}\relax
\EndOfBibitem
\bibitem[{de Bruijn}(1981)]{deBruijn1981pt2}
N.~{de Bruijn}, \emph{Indagationes Math.}, 1981, \textbf{84}, 53--66\relax
\mciteBstWouldAddEndPuncttrue
\mciteSetBstMidEndSepPunct{\mcitedefaultmidpunct}
{\mcitedefaultendpunct}{\mcitedefaultseppunct}\relax
\EndOfBibitem
\bibitem[G{\"a}hler(1988)]{Gahler1988}
F.~G{\"a}hler, Quasicrystalline Materials Proceedings, 1988, p. 272–284\relax
\mciteBstWouldAddEndPuncttrue
\mciteSetBstMidEndSepPunct{\mcitedefaultmidpunct}
{\mcitedefaultendpunct}{\mcitedefaultseppunct}\relax
\EndOfBibitem
\bibitem[Kiselev \emph{et~al.}(2012)Kiselev, Engel, and Trebin]{Kiselev_2012}
A.~Kiselev, M.~Engel and H.-R. Trebin, \emph{Phys. Rev. Lett.}, 2012,
  \textbf{109}, 225502\relax
\mciteBstWouldAddEndPuncttrue
\mciteSetBstMidEndSepPunct{\mcitedefaultmidpunct}
{\mcitedefaultendpunct}{\mcitedefaultseppunct}\relax
\EndOfBibitem
\bibitem[Glaser \emph{et~al.}(2015)Glaser, Nguyen, Anderson, Lui, Spiga,
  Millan, Morse, and Glotzer]{hoomd_2015}
J.~Glaser, T.~D. Nguyen, J.~A. Anderson, P.~Lui, F.~Spiga, J.~A. Millan, D.~C.
  Morse and S.~C. Glotzer, \emph{Comput. Phys. Commun.}, 2015, \textbf{192},
  97--107\relax
\mciteBstWouldAddEndPuncttrue
\mciteSetBstMidEndSepPunct{\mcitedefaultmidpunct}
{\mcitedefaultendpunct}{\mcitedefaultseppunct}\relax
\EndOfBibitem
\bibitem[Anderson \emph{et~al.}(2008)Anderson, Lorenz, and
  Travesset]{hoomd_2008}
J.~A. Anderson, C.~D. Lorenz and A.~Travesset, \emph{J. Comput. Phys.}, 2008,
  \textbf{227}, 5342--5359\relax
\mciteBstWouldAddEndPuncttrue
\mciteSetBstMidEndSepPunct{\mcitedefaultmidpunct}
{\mcitedefaultendpunct}{\mcitedefaultseppunct}\relax
\EndOfBibitem
\bibitem[Frenkel and Ladd(1984)]{Frenkel_1984}
D.~Frenkel and A.~J.~C. Ladd, \emph{J. Chem. Phys.}, 1984, \textbf{81},
  3188--3193\relax
\mciteBstWouldAddEndPuncttrue
\mciteSetBstMidEndSepPunct{\mcitedefaultmidpunct}
{\mcitedefaultendpunct}{\mcitedefaultseppunct}\relax
\EndOfBibitem
\bibitem[Addula \emph{et~al.}(2021)Addula, Veesam, and
  Punnathanam]{Punnathanam2021}
R.~K.~R. Addula, S.~K. Veesam and S.~N. Punnathanam, \emph{Molecular
  Simulation}, 2021, \textbf{47}, 824--830\relax
\mciteBstWouldAddEndPuncttrue
\mciteSetBstMidEndSepPunct{\mcitedefaultmidpunct}
{\mcitedefaultendpunct}{\mcitedefaultseppunct}\relax
\EndOfBibitem
\bibitem[Mermin(1968)]{Mermin_1968}
N.~D. Mermin, \emph{Phys. Rev.}, 1968, \textbf{176}, 250--254\relax
\mciteBstWouldAddEndPuncttrue
\mciteSetBstMidEndSepPunct{\mcitedefaultmidpunct}
{\mcitedefaultendpunct}{\mcitedefaultseppunct}\relax
\EndOfBibitem
\bibitem[van Smaalen(2007)]{Sander2007}
S.~van Smaalen, \emph{Incommensurate Crystallography (International Union of
  Crystallography Monographs on Crystallography)}, Oxford University Press,
  Oxford, 2007\relax
\mciteBstWouldAddEndPuncttrue
\mciteSetBstMidEndSepPunct{\mcitedefaultmidpunct}
{\mcitedefaultendpunct}{\mcitedefaultseppunct}\relax
\EndOfBibitem
\bibitem[Bak(1982)]{PBak_1982}
P.~Bak, \emph{Rep. Prog. Phys.}, 1982, \textbf{45}, 587\relax
\mciteBstWouldAddEndPuncttrue
\mciteSetBstMidEndSepPunct{\mcitedefaultmidpunct}
{\mcitedefaultendpunct}{\mcitedefaultseppunct}\relax
\EndOfBibitem
\bibitem[Bak(1986)]{PBak_1986}
P.~Bak, \emph{Physics Today}, 1986, \textbf{39}, 38--45\relax
\mciteBstWouldAddEndPuncttrue
\mciteSetBstMidEndSepPunct{\mcitedefaultmidpunct}
{\mcitedefaultendpunct}{\mcitedefaultseppunct}\relax
\EndOfBibitem
\end{mcitethebibliography}
\bibliographystyle{rsc} 
\end{document}